%% file: FairLedger.tex
\documentclass[a4paper]{article}[11pt]
\usepackage[sort]{cite}

\usepackage{algorithmicx}
\usepackage{algorithm}
\usepackage[noend]{algpseudocode}
\usepackage{amsthm}
\usepackage{amssymb}
\usepackage{amsmath}
\usepackage{tcolorbox}
\usepackage{graphicx}
\usepackage{caption}
\usepackage{subcaption}
\usepackage[margin=1in]{geometry}

\algnewcommand{\LeftComment}[1]{\Statex \(\triangleright\)
#1}

\algdef{SN}[LOCAL]{Local}{EndLocal}[1]{
\hspace*{-0.2cm}\textbf{Local state:}\ #1\ }{}%

\newcommand\sasha[1]{{\color{blue}[Sasha: #1]}}
\newcommand{\omitit}[1]{}

\algdef{SE}[RECEIVING]{Receiving}{EndReceiving}[1]{\textbf{upon
receiving}\ #1\ \algorithmicdo}{\algorithmicend\ \textbf{}}%
\algtext*{EndReceiving}

\algdef{SE}[UPON]{Upon}{EndUpon}[1]{\textbf{upon
}\ #1\ \algorithmicdo}{\algorithmicend\ \textbf{}}%
\algtext*{EndEvent}

\algblockdefx[OPERATION]{Operation}{EndOperation}[1]
  {{\bf operation} #1\   \algorithmicdo }
  {\algorithmicend\ }
\makeatletter
\ifthenelse{\equal{\ALG@noend}{t}}%
  {\algtext*{EndOperation}}
  {}%

\theoremstyle{definition}
\newtheorem{definition}{Definition}

\usepackage{datetime}
\usepackage{url}
\usepackage{hyperref}

\date{}

\begin{document}

\title{FairLedger: A Fair Blockchain Protocol for
Financial Institutions}

\author{
Kfir Lev-Ari\thanks{Kfir Lev-Ari and Alexander Spiegelman contributed
equally to this work.}\\
Technion IIT
\and Alexander Spiegelman\footnotemark[1]\\
VMware Research
\and Idit Keidar\\
Technion IIT
\and Dahlia Malkhi\\
VMware Research}


\maketitle

\input{sections/abstract.tex}

\input{sections/introduction.tex}

\input{sections/structure.tex}

\newpage
\bibliographystyle{acm}
\bibliography{bibliography}


\end{document}

%% file: sections/abstract.tex
%

\begin{abstract}
Financial institutions are currently looking into
technologies for permissioned blockchains.
A major effort in this direction is Hyperledger, an
open source project hosted by the Linux Foundation and
backed by a consortium of over a hundred companies.
A key component in permissioned blockchain protocols is a
byzantine fault tolerant (BFT) consensus engine that
orders transactions.
However, currently available BFT solutions in Hyperledger
(as well as in the literature at large) are
inadequate for financial settings; they are not
designed to ensure fairness or to
tolerate selfish behavior that arises when financial
institutions strive to maximize their own profit.

We present FairLedger, a permissioned blockchain BFT
protocol, which is fair, deigned to deal with rational
behavior, and, no less important, easy to understand and
implement.
The secret sauce of our protocol is a new communication
abstraction, called detectable all-to-all (DA2A), which
allows us to detect participants (byzantine or
rational) that deviate from the protocol, and punish
them.
We implement FairLedger in the Hyperledger open
source project, using Iroha framework, 
one of the biggest projects therein.
To evaluate FairLegder's performance, we also
implement it in the PBFT framework and compare the two
protocols.
Our results show that in failure-free scenarios FairLedger
achieves better throughput than both Iroha's 
implementation and PBFT in wide-area settings.

\end{abstract}

%% file: sections/introduction.tex
\section{Introduction}
\label{introduction}


As of today, support for financial transactions between
institutions is limited, slow, and costly. 
For example, an oversees money transfer between two banks might
take several days and entail fees of tens of dollars.
The source of this cost (in term of both time and money)
is the need for a reliable clearing house; sometimes this even
requires physical phone calls at the end of the day to make sure
all the balances coincide.
At the same time, emerging decentralized 
cryptocurrencies like
Bitcoin~\cite{nakamoto2008bitcoin} complete transactions
within less than hour, at a cost of microcents.
It is therefore not surprising that financial institutions
are looking into newer technologies to bring them up to
speed and facilitate trading in today's global economy.

Perhaps the most prominent technology considered in this
context is that of a \emph{blockchain},
which implements a secure peer-to-peer
\emph{ledger} of financial transactions on top of a
consensus engine.
A major effort in this direction is
Hyperledger~\cite{hyperledger}, an open source project
hosted by the Linux Foundation and backed by a consortium of more than a hundred companies. 
In contrast to cryptocurrency protocols deployed over the
Internet, which are fully anonymous and allow any party to join
or leave at any time, blockchain protocols for financial
institutions, also called \emph{permissioned blockchains}, are
much more conservative:
Every participant is known and certified, so that it has
to be responsible for its actions in the real world.
In addition, such systems are intended to be deployed over
a secure and reliable wide-area network (WAN).
Therefore, proposed solutions for permissioned
blockchains~\cite{miller2016honey, iroha, hyperledger}
abandon the slow and energy-consuming proof-of-work
paradigm of Bitcoin, and tend to go back to more
traditional distributed consensus protocols.
Because of the high stakes, malicious deviations from the
protocol (due to bugs or attacks), as
rare as they might be, should never compromise the
service.
Such deviations are modeled as \emph{byzantine}
faults~\cite{lamport1982byzantine}, and to deal with them,
proposed solutions use \emph{byzantine fault tolerant (BFT)} consensus
protocols.

Yet, dealing with byzantine failures is only a
small part of what is required in permissioned
blockchains.
In fact, a break-in that causes a bank's software to
behave maliciously is so unusual that it is a top news story,
and  is investigated by official authorities such as
the FBI.
On the other hand, financial institutions always
try to maximize their own profit, and would never use a
system that discriminates against them.
Moreover, they can be expected to 
selfishly deviate from the protocol whenever they can
benefit from doing so.
In particular, financial entities typically receive a
fee for every transaction they append to the ledger, and
thus can be expected to attempt to game the system in a
way that to maximizes the rate of their transactions in
the ledger.
Such \emph{rational} behavior, if not carefully
considered, not only can discriminate against some of
the entities, but may also compromise safety.


As a result, in the FinTec context, one faces a number of
important challenges that were not emphasized in
previous BFT work:
(1) \emph{fairness} in terms of the opportunities each
participant gets to append transactions to the ledger;
(2) expected \emph{rational behavior} by all parties;
and (3) \emph{optimized failure-free performance}, given
that financial institutions are usually very secure.
In addition, it is important to stress 
(4) \emph{protocol simplicity}, because complex protocols
are inherently bug-prone and easier to attack.
In this work we develop FairLedger, a new BFT
permissioned blockchain protocol for the Hyperledger
framework, which addresses all of these challenges.
Our protocol is fair, designed for rational participants,
optimized for the failure-free case,
simple to understand, and easy to implement.
Specifically, we show that following the protocol is an
equilibrium, and when rational participants do follow the
protocol, they all get perfectly fair shares of the ledger.

Given that byzantine failures are expected to be rare, our
philosophy is to optimize for the ``normal mode'' when
they do not occur (as also emphasized in some previous work,
e.g., Zyzzyva~\cite{kotla2007zyzzyva}).
For this mode, we design a simple protocol that
provides high performance when all parties are rational
but not byzantine. 
Under byzantine failures, the normal mode protocol
remains safe and fair, but does not necessarily guarantee
progress.
Upon detecting that a rogue participant is
attempting to prevent progress, we switch to the
``alert mode".
At this point, it is expected that real-world
authorities (such as the FBI or Interpol) will step in
and investigate the break-in. But such an investigation
may take days to complete, and in the time being, the
service remains operational using the alert mode
protocol, even if at degraded performance.

An important lesson learned from the deployment of
Paxos-like protocols in real systems such as
ZooKeeper~\cite{hunt2010zookeeper} and etcd~\cite{etcd}, 
is that systems will only be used if they are easy to understand, implement, and maintain. 
Specifically, Paxos-like protocols use a quorum to agree
on every transaction appended to the ledger. 
Whereas the general Paxos~\cite{lamport2001paxos} allows
using a different quorum for every transaction, practical
systems do away with this freedom.
Instead, they follow the Vertical
Paxos~\cite{lamport2009vertical, abraham2016bvp} approach
of using a fixed quorum for a sequence of transactions, and
reconfiguring it upon failures~\cite{hunt2010zookeeper,
etcd}.
We follow this approach in FairLedger. 
Specifically, we designate a \emph{committee} of participants
who are interested in issuing transactions (e.g., banks)
and have them run a \emph{sequencing protocol} to order
all their transactions.
A complementary \emph{master} service
monitors the committee's progress and initiates
reconfiguration when needed.
Including all interested parties in quorums is also
instrumental in achieving fairness-- this way, all the
committee members benefit from sequencing batches that
include transactions by all of them. 
We use rate adjustments, batching, and asynchronous
broadcast to achieve high throughput even if some
committee members are slow.

In the absence of failures, the committee runs an
efficient normal mode sequencing protocol.
In this mode, byzantine participants cannot violate safety but
may prevent progress, causing the master to switch
the system to the alert mode.
We assume a loosely synchronous model, where a
master can use a coarse time bound (e.g., one minute) to
detect lack of progress.
This bound is only used for failure recovery, and does not
otherwise affect performance.  
The key feature of our alert mode sequencing protocol is that if
participants deviate from the protocol in a way that jeopardizes
progress, they are accurately detected.
If they are slow, their transaction rate is lowered, and if
they are not cooperative, they are removed from the
committee altogether.
Unlike in other Hyperledger protocols~\cite{iroha},
FairLedger never indicts correct participants. 
Identifying faulty
components without accusing correct ones is essential in
allowing the system to heal itself following attacks.

The sequencing protocol uses all-to-all communication
among committee members.
Since the quorum includes all participants and all
messages are signed, the protocol ensures safety despite
byzantine failures of almost any minority.
Specifically, for $f$ failures, our protocol is correct when the
number of participants $n \geq 2f+3$.

Nevertheless, it is enough for
one participant to withhold a single
message in order to prevent progress.
Such a deviation from the protocol is hard to
detect even with reliable communication since one
participant can claim that it sent a message to another,
while the recipient claims the message was not sent. 
To deal with such deviations we define
a new communication abstraction, which we call
\emph{detectable all-to-all (DA2A)}.
Besides the standard \emph{broadcast} and
\emph{deliver} API, DA2A exposes a
\emph{detect} method that returns an accurate set of
participants that deviated from the broadcast or deliver
protocols.

We implement FairLedger's sequencing protocol in
Iroha~\cite{iroha}, which is part of the
Hyperledger~\cite{hyperledger} open-source project, and compare
its performance to their up-to-date version.
Specifically, since Iroha's implementation is modular, we
are able to replace their BFT consensus protocl (which is based
on~\cite{duan2014bchain}) with our sequencing protocol without
changing other components (e.g., communication,
cryptographic, and database libraries). 
We compare both versions in Emulab~\cite{emulab},
configured to simulate a wide-area network, and our results show
that FairLeadger outperforms Iroha's BFT protocol in the vast
majority of the tested scenarios (both in normal mode and in 
alert mode).

In addition, since the Iroha system consists of many
components (e.g., GRPC~\cite{grpc} communication) that may
include overheads and bottlenecks, we also
implement FairLedger's sequencing protocol in the
PBFT~\cite{pbftCode} framework, which provides a
clean environment to check performance.
Again, we use Emulab~\cite{emulab}, configured to
simulate a wide-area network, to compare FairLedger to the steady
state protocol of PBFT~\cite{castro1999practical}.
Our results show that Fairledger's latency is better than PBFT's
in both the normal and alert modes. 
Fairledger's throughput
outperforms PBFT's in normal mode and is inferior to it in the
alert mode, but PBFT's advantage is diminished as the system
scale grows.

In summary, this paper makes the following contributions:

\begin{enumerate}
  
  \item We define a fair distributed ledger abstraction
  for rational participants.
  
   \item We define a detectable all-to-all
  (DA2A) abstraction that identifies
  participants deviating from the communication protocol. 
  
  \item We design FairLedger, the first BFT blockchain protocol
  that ensures strong fairness when all participants are rational:
  FairLedger is safe under byzantine failures of almost any
  minority, and detects and punishes deviating (byzantine and
  rational) participants.
  It is also simple to understand and implement.

  \item We implement FairLedger's sequencing protocol in the
  Hyperledger framework and substitute it for Iroha, a ledger 
  solution included therein.
  Our results show that FairLedger outperforms Iroha's original
  BFT implementation in the vast majority of cases.
  
  \item We implement and test FairLedger's sequencing protocol in
  the PBFT framework.
  Our result shows that FairLedger outperforms PBFT in the
  normal mode, and achieves slightly lower results in the alert
  mode.

\end{enumerate}

The rest of the paper is organized as follows:
Section~\ref{sec:Ledger} defines rational participants
and the fair ledger service, while
Section~\ref{sec:sesmodel} details our system model.
In Section~\ref{sec:sesarch} we present our
architecture, and in Section~\ref{sec:protocol} we
give the FairLedger protocol.
In Section~\ref{sec:imp}, we describe the implementation
in Hyperledger and in the PBFT frameworks, and in
Section~\ref{sec:eval} we evaluate our protocol in the
both of them.
Finally Section~\ref{sec:related} discusses related
work, and Section~\ref{sec:discussion} concludes the
paper.

%% file: sections/structure.tex
\section{A Fair Ledger Abstraction for Rational Players}
\label{sec:Ledger}

We consider a set of players, each of which represents a
real-world financial entity (e.g., a bank), jointly
attempting to agree on a shared \emph{ledger} of financial transactions. 
Every player has an unbounded stream of transactions that it
wants to append to the common ledger.
The stream, for example, can come from the entity's clients.
We assume that the financial entity receives a fee (or some
other benefit) for every transaction it appends to the shared
ledger, and so it is motivated to append as many transactions as
possible.
A principal goal for our service is \emph{fairness}, that
is, providing entities with equal opportunity for
appending transactions.
Section~\ref{sub:BAR} discusses how we model the players, and 
Section~\ref{sub:ledger} defines the fair shared ledger
abstraction.

\subsection{Byzantine and rational behavior}
\label{sub:BAR}

Traditional distributed systems are usually managed by one
organization, and thus whenever an entity deviates from the
protocol, it can be explained as a software or hardware bug or
by this entity being hacked.
Therefore, protocols for such environments
are designed to remain correct even if some entities
deviate from the protocol in an arbitrary manner.
Such protocols are called \emph{byzantine fault tolerant}, and
obviously only a small subset of the entities are allowed
to be byzantine.
But, since in this work we seek a protocol that
coordinates among many organizations, and especially because financial assets are
involved, we have to take into account that \emph{every} entity
may behave \emph{rationally}, and deviate from the protocol if
doing so increases its benefit.

To reason about such rational behavior we follow~\cite{Moscibroda},
and assume that each entity can be either \emph{byzantine} or
\emph{rational}.
A rational entity has a known utility function that it
tries to maximize and deviates from the protocol only if this
increases its utility, whereas a byzantine entity can deviate
arbitrarily from the protocol,i,e, its utility function is unknown. 


We assume that the system involves two types of entities --
\emph{players} and \emph{auditors}. 
Players (e.g., banks) propose transactions they would like to
append to the ledger, while auditors oversee the system. 
The same physical entity may be both a player and an auditor,
but additional entities (e.g., government central banks) may act
as auditors as well. There are initialy $n$ players, and any number
of auditors.
The number of byzantine players is bounded by a known parameter
$f$, and we require $n \geq 2f+3$. 
In addition, at most a minority of the auditors can
be byzantine. 
We assume that byzantine entities can collude, but rational
ones do not.

In order to prove that a protocol is correct in our
model, we need to show that (1) the problem specification is
satisfied in case all the rational entities follow the protocol
and there are at most $f$ byzantine ones, and (2) following
the protocol is an equilibrium
for rational entities even in the presence of $f$ byzantine ones.
These two conditions imply the protocol's correctness assuming
that players do not deviate unless they benefit from doing so. 
A similar assumption was made in previous works on BAR (byzantine,
altruistic, rational) fault tolerance~\cite{aiyer2005bar, li2006bar}.
 
\subsection{Distributed fair ledger}
\label{sub:ledger}

A \emph{log} is a sequence of \emph{transactions} from some
domain $\mathcal{T}$.
Note that $\mathcal{T}$ is defined by the high-level application
that uses our ledger abstraction. 
A \emph{ledger} is an abstract object that maintains a log
(initially empty) and supports two operations, \emph{append} and
\emph{read}, with the following sequential specification: 
An \emph{append$(t)$} operation with $t \in \mathcal{T}$ changes
the state of the log by appending $t$ to the end of the log. 
A \emph{read$(l)$} operation returns the last $l$
transactions in the log.

The \emph{utility function of a rational player} is the
ratio of transactions that it appends to the ledger, i.e., the number
of transactions it appends to the ledger out of the total number of
transactions in the ledger. If two ledgers has the same ratio, then
the one with the more transactions is preferred.
Meaning that the players care about the overall system progress but
they care more about getting a fair share of it.
%
%

The \emph{utility function of an auditor} is the following: an
auditor that is also a player has the player's utility function.
Otherwise, its utility function is the number of players on the
committee in case progress is being made, and 0 in case the system
stalls.
In other words, the auditors aim is to ensure the system's overall
health, which mean not to remove a player unless it couses the system
to stall.
%

We enforce strict fairness. 
Intuitively, this means that every player gets an equal
number of opportunities to append a transaction to the log.
Thus, if player $p_1$ follows the protocol, then at
any point when the log contains $k$ transactions appended by
$p_1$, the log does not contain more than $k+1$
transactions appended by any other player.
In Section~\ref{sec:sesmodel} we formalize and extend this
definition to a case in which different players are allocated
different shares of the log, and these shares (as well
as the set of players) may change over time.

A \emph{distributed ledger} protocol emulates an abstract
ledger with atomic operations to a set of players that access it
concurrently.
The shared ledger state at any time 
(1) reflects all completed operations by players that
follow the protocol, and (2) may or may not reflect
pending (not completed) operations as well as operations
performed by players (byzantine or rational) that deviate
from the protocol.
Intuitively, this means that the protocol tolerates
players that do not cooperate by restricting the possible
outcomes of their behavior to executing correct
operations or leaving the log's state unchanged.

\section{System Model}
\label{sec:sesmodel}

We now state our assumptions on the deployment environment of our
protocol.

\paragraph{Certificates.}
We assume that players have been certified by some
trusted certification authority (CA) known to all players.
In addition, we assume a PKI~\cite{rivest1978method}:
each player has a unique pair of public and private cryptographic keys, where the public
keys are known to all players, and no coalition of players has
enough computational power to unravel other players' private
keys.

\paragraph{Reliable communication.} We assume reliable
communication channels (implemented, e.g., using TCP or
using retransmissions over UDP) between pairs of players.
Such channels are not strictly required among all pairs,
but there must be at least $2f+1$ players that communicate
reliably with all others.

\paragraph{Timing assumptions.}
As in previous works on permissioned blockchains~\cite{iroha,
duan2014bchain, hyperledger}, we assume that there is a known
upper bound $\Delta$ on message latency.
Nevertheless, our sequencing protocol is safe and fair even if
the bound does not hold. 
We exploit this bound to detect failures when the
protocol stalls because a byzantine (or rational) player
deviates from the protocol by withholding messages.
Thus, the bound can  be set very conservatively (e.g., in
the order of minutes) so as to avoid false detection.


\paragraph{Rational and byzantine behavior.}

We assume that rational entities do not collude, but byzantine
players are controlled by a strong adversary, and thus can
arbitrarily deviate from the protocol (e.g., crash, withhold
messages, or send incorrect protocol messages) and collude.
Because we assume synchrony and a PKI, we can
overcome byzantine failures of almost any
minority~\cite{dolev1983authenticated}.

\paragraph{Quality of service.}

Above, we gave a simplistic definition of fairness assuming all
players are allowed to append transactions at the same rate.
However, this does not necessarily have to be the case.
For example, slow players sometimes cannot sustain the throughput
required by others, and thus by insisting on strict fairness we
decrease the total system throughput.
In addition, it is possible that some players deserve more
throughput, e.g., because they pay more for the service.
Therefore, we generalize our fairness definition to
allow general quality of service (QoS) allocations.
Because QoS allocations may change over time, we define
\emph{QoS-fairness} for segments of the ledger.
Denote by $TX[i,j]$ the segment of the ledger from the i$^{th}$
to the j$^{th}$ entry (inclusive).

\begin{definition}[QoS]

Given a tuple $R= \langle r_1,r_2,\ldots,r_n \rangle$
s.t.\ $\forall i~ 0\leq r_i \leq 1$ and $\Sigma_{i=1}^n r_i =1$,
we say that the segment $TX[i,j]$ of a (sequential) ledger is
\emph{R-fair} if for every player $p_i$ that follows the
protocol, the number of transactions in $TX[i,j]$ that were
appended by $p_i$ is at least $\lfloor |TX[i,j]| r_i
\rfloor$. 

\end{definition}

\noindent Note that the ledger fairness definition
from the previous section coincides with the \emph{R-fair}
ledger definition when for every $r_i \in R$, $r_i = 1 / n$.

\section{System Architecture}
\label{sec:sesarch}



Our goal in this paper is to design a ledger
protocol that financial institutions will be able to use.
Such a protocol, besides being fair, secure against
malicious attacks, and resilient to selfish behavior,
must be simple to understand, implement, and maintain.
Therefore, although we appreciate, from both theoretical
and practical perspectives, complex protocols with many
corner cases and clever optimizations, we try here to
keep the design as simple as possible.
The simple design not only reduces vulnerabilities, it
also makes it much easier to reason about selfish
behavior.

\begin{figure}[H]      
  \centering 
  \includegraphics[width=3in]{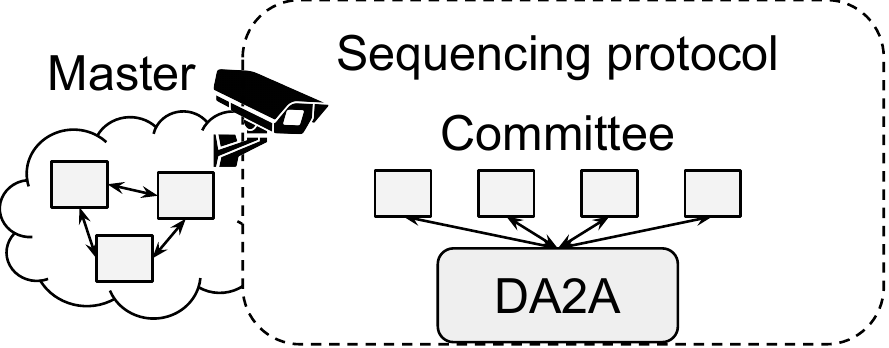}
  \caption{System architecture. The sequencing protocol is run
  by a committee using a detectable all-to-all (DA2A) service
  and is monitored by a master.}
 
   \label{fig:system}
\end{figure}

\paragraph{Committee and master.}
We adopt the Vertical Paxos~\cite{lamport2009vertical,
abraham2016bvp} paradigm, which unlike the original Paxos
protocol does not allow different quorums to be used for
different transactions.
Instead, there is a single (known to all) quorum, called
\emph{committee}, which partakes in agreeing on every
transaction.
Initially, the committee consists of all players.
By requiring all committee members to endorse
transactions, we create an incentive for all of them to
sequence batches of transactions from all of them. 
To handle cases when committee members stop responding
(e.g., due to a crash or an attack), a complementary
\emph{master} service performs \emph{reconfiguration}: detecting
such members and removing (or replacing) them.
Thus, we logically implement two components: (1) a
committee of $n$ players that runs the sequencing
protocol to append transactions to the ledger, and (2) a
master, which is responsible for progress and determines
the QoS allocations; see Figure~\ref{fig:system}.
The master is implemented by auditors using a minority-resilient
synchronous BFT protocol~\cite{dolev1983authenticated}.
Its impact on overall system performance is small, and so we do
not focus on distributed implementation of the master in this
paper.
For our purposes, the master is a single trusted authority.

The committee's sequencing protocol is
implemented on top of a communication primitive we call
\emph{detectable all-2-all (DA2A)}, as explained shortly.
The sequencing protocol is safe and fair even if almost any
minority of players are byzantine, but may stall even if
only one player deviates from the protocol


The role of the master is to monitor the protocol and detect
players that prevent progress.
When lack of progress is observed, the master runs a
\emph{recovery} protocol to reconfigure the system: it
removes byzantine players, possibly adds new players, and
punishes slow or selfish players by reducing their ratio
in the ledger (i.e., changes the QoS allocation).

\paragraph{QoS adjustment.}
In addition to forming the committee, the master also determines
the QoS that should be enforced by it.
Every time the master reconfigures the system, it provides a new
vector $R = \langle r_1,\ldots, r_n \rangle$ that represents the
ratio each committee member should get in the ledger.
%
%
The portion of the log decided by the new committee satisfies the
\emph{QoS-fairness} requirement with respect to $R$.

The initial value of the QoS allocations can be chosen
based on real-world contracts among the financial
institutions, or by their available throughput or payment. 
Subsequently, the master's authority to modify the
QoS enforced by the protocol empowers it both to ensure
that rational players follow the protocol and to adjust
the bandwidth allocations to player capabilities:
Whenever the master detects a player that sends messages
at a low rate or deviates from the protocol, it
immediately reduces the ratio of transactions that player
gets.
A rational player, whose utility function is the ratio
of transactions it appends to the log, will prefer to
collaborate in fear of such punishment.

\paragraph{Detectable byzantine broadcast.}
The master's ability to use the punishment mechanism as
well as to evict byzantine players relies on its ability
to detect deviations from the protocol.
We divide the possible deviations into two categories:
\emph{active} and \emph{passive}.
An active deviation occurs when a player tries to break
consistency or fairness by sending messages that do not coincide with the
protocol.
By singing all messages with private keys, we achieve
non-repudiation, i.e., messages can be linked to their
senders and provide evidence of misbehavior, which the master
can use to detect deviation from the protocol.

Passive deviation, which stalls the protocol by withholding
messages, is much harder to detect.
Even a single player can stop our sequencing protocol's
progress by simply not sending messages.
Accurately detecting passive deviation is impossible in
asynchronous systems, and is not an easy
task even in synchronous systems with reliable
communication.
For example, if the protocol hangs waiting for $p_1$ to take an
action following a message it expects from $p_2$, we
cannot, in general, know if $p_2$ is the culprit (because
it never sent a message to $p_1$) or $p_1$ is at fault.

To address this problem we present a new
broadcast abstraction, which we call \emph{detectable
all-to-all (DA2A)}, and use it  in our
sequencing protocol.

\begin{definition}[DA2A]

Consider $n$ players and a master.
The API of \emph{DA2A} supports \emph{broadcast$(m)$} and
\emph{deliver(m)} operations for the players, and a
\emph{detect()} operation for the master.
Every player $p_i$ invokes \emph{broadcast$(m_i)$} for some
message $m_i$ s.t.\ all the other players should
\emph{deliver$(m_i)$}.
The \emph{detect()} operation performed by the master
returns a set $S$ of players that deviate from
the protocol together with corresponding proofs; for
every two players $p_j,p_i$ s.t.\ $p_i$ does not deliver
a message from $p_j$, $S$ contains $p_j$ (with a proof of
$p_j$'s deviation) in case $p_j$ did not perform
$broadcast(m)$ properly, and otherwise, it contains $p_i$
(with a proof of $p_i$'s deviation).

\end{definition}

Note that in case $S$ is empty, all the players follow
the protocol, meaning that all the players broadcast a
message and deliver messages broadcast by all other
players.
Clearly, implementing DA2A, and in particular, its
$detect()$ method requires an upper bound on message latency and
a correct majority.
We present an implementation under this assumptions in 
the next section.

\section{FairLedger Protocol}
\label{sec:protocol}

We start by presenting our detectable all-to-all building
block in Section~\ref{seb:DA2A}.
Then, we describe how we use it for our sequencing protocol in
Section~\ref{sub:SS}, and for the recovery protocol in
Section~\ref{sub:rec}.
Finally, in Section~\ref{sub:proof}, we give correctness
arguments.

\subsection{Detectable all-to-all (DA2A)}
\label{seb:DA2A}

\begin{figure}[t!]
    \centering
    \begin{subfigure}[t]{0.2\textwidth}
        \centering
        \includegraphics[height=1.5in]{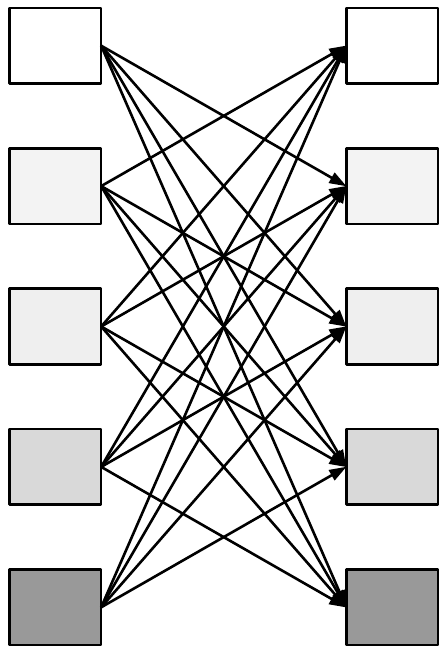}
        \caption{direct all-to-all}
        \label{fig:sub:r1}
    \end{subfigure}%
    ~ 
    \begin{subfigure}[t]{0.3\textwidth}
        \centering
        \includegraphics[height=1.5in]{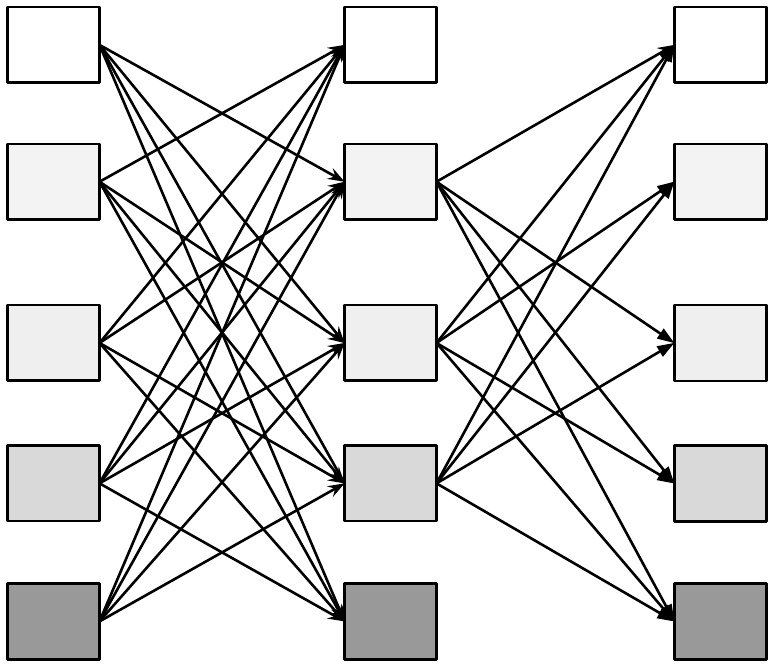}
        \caption{relayed all-to-all}
        \label{fig:sub:r2}
    \end{subfigure}
    \caption{All-to-all communication patterns.}
\end{figure}

\paragraph{Communication patterns.}
We start by discussing two ways to implement
all-to-all communication over reliable links.
The simplest way to do so is \emph{direct all-to-all},
in which \emph{broadcast(m)} simply sends message $m$ to
all other players (see Figure~\ref{fig:sub:r1}).
This implementation has the optimal cost of 1 hop and
$n^2$ messages, but cannot reveal any information about
passive deviations: In case $p_i$ does not deliver any
message from $p_j$, the master has no way of knowing
whether $p_j$ did not send a message to $p_i$, or $p_i$
is lying about not receiving the message.

Another way of implementing all-to-all communication is
by using a subset of the players as \emph{relays}.
We call this approach \emph{relayed all-to-all}.
In this approach, a \emph{broadcast($m$)} operation
sends $m$ to all the players, and when a relay receives a
message for the first time, it forwards it to all players
(see Figure~\ref{fig:sub:r2}).
For $r$ relays this requires 1-2 hops depending whether
any of the players are byzantine and $n(n+rn) = (r+1)n^2$
messages.


Note that when using relays, it is possible to have
players send their messages only to the relays.
This induces lower overhead but takes longer in case all
players cooperate. 
This approach may be used when direct all-to-all communication
is not feasible. 
For example, in case the system is deployed on top of
private physical links, such links might not necessarily exist among all pairs of players.
Similarly, note that the relayed communication does not
necessarily have worse latency than direct all-to-all,
since the latter depends on the slowest link, while in
relayed communication we can pick the relays with the
fastest links.

\paragraph{Detectability.}
Obviously, we cannot implement detectable all-to-all
(DA2A) broadcast with direct all-to-all communication.
On the other hand, if we assume that all non-byzantine players follow
the protocol, we can perfectly detect a guilty party
by using $2f+1$ relays.
A \emph{detect$()$} operation by the master waits
$2\Delta$ time to make sure that all messages that were
sent arrived, and then for every two players $p_j$ and
$p_i$ s.t.\ $p_i$ does not deliver a message from $p_j$,
it asks the relays whether they received a message from
$p_j$.
The relays' replies are signed and used as proof of a
deviation.
In case $f+1$ relays say yes, then at least one correct
relay received a message from $p_j$ and sent it to $p_i$,
meaning that $p_i$ received it -- recall that we assume
reliable communication -- and deviated from the protocol
by not delivering it.
Otherwise, $p_j$ did not send a message to all
relays, meaning that $p_j$ deviated from the broadcast
protocol.
Thus, either way, the master can detect the faulty
player.
It is important to notice that the detection is accurate:
it has no false positives, and finds the player
responsible for every message omission.

In order to prove that following the protocol is a Nash equilibrium
for rational player, we need the $detect()$
method to tolerate one more possible deviation by a non-byzantine
player; that is, we need to accurately detect passive deviations that
stall progress even if $f+1$ players deviate from the protocol.
Note that when a progress problem is caused by a
player $p_i$ failing to deliver a message broadcast by
player $p_j$, we know that at least one of them deviates
from the protocol. 
Thus, at most $f$ of the remaining players may deviate.
Therefore, it is enough to pick $2f+1$
players different from $p_i$ and $p_j$ to be the
relays in order to identify the culprit in case
the problem is not solved.
Note that this is always possible since we assume  $n \geq 2f+3$.

\paragraph{Practical deployment.}

Since we assume byzantine failures are very rare, a
practical strategy is to employ direct all-to-all
communication (in case it is feasible) for as long as
there is progress. 
In case direct communication among players is unavailable
or slow, we can use any number of relays (e.g., one
relay). 
We call this the normal mode. 
In case of a progress problem, we switch to the degraded
alert mode, with $2f+1$ relays.
If the progress problem is not resolved, it is
guaranteed that the master detects the misbehaving
players, and replaces them.
At this point, we can switch the
system back to the normal mode.
Note, however, that byzantine players can avoid
detection by behaving properly in the alert mode.
They can thus force the system to stay in this mode and
continue to send more messages (by using relays), but do
not compromise progress.
In the meantime, an official external authority (e.g., FBI
or Interpol) can investigate the security breach to find
the misbehaving component.

\subsection{Sequencing protocol}
\label{sub:SS}


The sequencing protocol works in \emph{epochs}, where
in each epoch every participating player gets an
opportunity to append one transaction or one
fixed-size batch of transactions to the log.
The key mechanism to ensure fairness is to commit all the
epoch's transactions to the log atomically (all or
nothing).
Since we assume that players have infinite streams of
transactions they wish to clear, they always have
enough transactions to append.
And if not, they can always append an empty (dummy)
transaction.

\begin{figure}[H]      
  \centering 
  \includegraphics[width=4in]{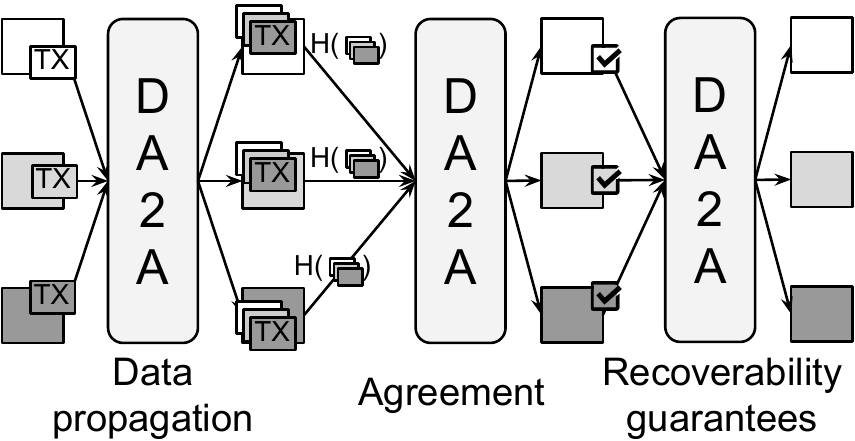}
  \caption{Sequencing protocol.}
 
   \label{fig:steadystate}
\end{figure}

An $append(t)$ operation locally buffers $t$ for
inclusion in an ensuing epoch, and waits for it to be
sequenced. 
Each epoch consists of three DA2A communication
rounds (see Figure~\ref{fig:steadystate}) among players
participating in the current epoch, proceeding as follows:


\begin{enumerate}
  
  \item Broadcast a transaction or batch to all; upon
  receiving transactions from all (including self), order
  received transactions by some deterministic rule and sign
  the hash $h$ of the sequence.
  
  \item Broadcast $h$ to all; receive from all and verify
  that all players signed the same hash.
   
  \item Broadcast $\langle commit, epoch, h \rangle$
  (signed) to all, return when receive the same message
  $f+1$ times.
 
\end{enumerate}

\noindent The sequencing protocol is described in
Algorithm~\ref{alg:player1}.
For clarity, we do not include signature manipulation
although all the messages are signed and verified; we
also present a version where the QoS allocation is
equal for all players.

\input{sections/playerCode1.tex}

The purpose of the first round is to broadcast
all the transactions of the epoch.
The second round ensures safety; at the end of this round 
each player validates that all other players signed the
same hash of transactions, meaning that only this hash can
be committed in the current epoch.
The last round ensures recoverability during
reconfiguration as we explain in Section~\ref{sub:rec}
below.
Note that we achieve fairness by waiting
for all players; an epoch is committed only if all
the players sign the same hash, and
since each player signs a hash that contains its own
transaction, we get that either all the players'
transactions appear in the epoch, or the epoch is not
committed.

\paragraph{Read operations.}
Since all players make progress together, they all have
up-to-date local copies of the ledger.
A $read(l)$ operation simply returns the last
$l$ committed transaction in the local ledger, where
for every returned sequence of transactions
$st$  pertaining to some epoch k, it attaches a proof for
$st$.
We need the attached proof in order to make sure byzantine
players do not lie about committed transactions. 
The proof is either (1) a $\mathsf{newConfig}$ message from the
master that includes $st$ (more details below), or (2) $f+1$
epoch $k$ round 3 messages, each of which contains a hash of
$st$.

\paragraph{Supporting quality of service.}
To support non-uniform quality of service we include
different batch sizes from different players in each
epoch.
For example, for players $p_1,p_2,p_3$, and vector
$\langle R=1/2,1/4,1/4 \rangle$, $p_1$ appends a batch of
two transactions in every epoch, whereas $p_2$ and $p_3$
each append one transaction.


\paragraph{Asynchronous broadcast.}

The first round of our sequencing protocol exchanges
transactions (data), the second round exchanges hashes of the
transactions (meta-data), and the last round exchanges commit
messages (meta-data).
Hence, the first round consumes most of the bandwidth.
In order to increase throughput, we decouple data
from meta-data and asynchronously broadcast transactions 
(i.e., execute the first round) of every epoch as soon as
possible.
However, in order to be able to validate transactions, we
perform rounds 2 and 3 sequentially.

In other words, we divide our communication into a data
path and a meta-data path, where the data path is
out-of-order and the meta-data path orders the data.
This is a common approach, used, for example, in atomic
broadcast algorithms that use reliable broadcast to
exchange messages and a consensus engine to order
them~\cite{cristian1986atomic, cachin2001secure}.

\subsection{Recovery}
\label{sub:rec}

Since we use a PKI, proving active deviations is easy, and
every time a player sees evidence for active deviation, it
sends it to the master.
One example appears in Algorithm~\ref{alg:player1}
line~\ref{line:complain}: a player gets two
different hashes (corresponding to different sequences of
transactions) in the second round, in which case, to
ensure correctness, it cannot move on to round three.
Instead, it complains to the master and waits for
reconfiguration.
When the master receives both hashes it checks which of
the players signed two different transactions in the
first round and issues a reconfiguration to remove that
player.
Other active deviations, e.g., incorrect messages
formats, are handled in a similar manner;
for simplicity we omit this from the code, and focus only
on processing of correctly-formatted messages.

The master's protocol is described in
Algorithm~\ref{alg:master}.
To detect passive deviations that prevent progress, we
use the $detect()$ operation exposed by the DA2A
abstraction.
The sequencing protocol is simply an infinite sequence of
DA2A instances.
Therefore, the master sequentially invokes $detect()$
(Algorithm~\ref{alg:master}, line~\ref{line:detect}) for all
epochs, until it returns a non-empty set $S$ indicating
that the sequencing protocol is stalled, in which case
the master invokes \emph{reconfigure$(S)$}
(Algorithm~\ref{alg:master},
lines~\ref{line:recB}-\ref{line:recE}).

First, it stops the current configuration and learns
its closing state by sending a $\mathsf{reconfig}$
message to the current committee.
To prove to the players on the committee that a
reconfiguration is indeed necessary, the master attaches
to the $\mathsf{reconfig}$ message proof reconfiguration
is warranted. 
this can be evidence of active
deviation, or a proof of passive deviation returned from
the $detect()$ method of the DA2A.
It can also be a real-world contract (signed by a CA) 
adding a new player or increasing a player's ratio. 
When a player receives a $\mathsf{reconfig}$ message
(Algorithm~\ref{alg:player1},
lines~\ref{line:reconfigBegin}-\ref{line:reconfigEnd}),
it validates the proof for the reconfiguration, sends its
local state (ledger) to the master, and waits for a
$\mathsf{newConfig}$ message from the master.
When a player receives $\mathsf{newConfig}$ with a new
configuration, it validates that every player addition or
remove is justified by a proof, and ignores 
requests that do not have a valid proof. 

\paragraph{State transfer.}
Note that while a byzantine player cannot make the master
believe that an uncommitted epoch was committed (a
committed epoch must be signed by all the epoch's
players), it can omit a committed epoch when asked (by
the master) about its local state.
Such behavior, if not addressed, could potentially lead
to a safety violation:
suppose that some byzantine player $p$ does not broadcast
its last message in the third round in epoch
$k$, but delivers messages from all other players.
In this case, $p$ has proof that epoch $k$ is
committed, and may return these transactions in response
to a read.
However, no other player has proof that epoch $k$ is
committed and $p$ withholds epoch k's commit from the
master.
In this case, the new configuration will commit different
transactions in epoch $k$, which will lead to a
safety violation when a \emph{read} operation will be
performed.

The third round of the epoch is used to overcome this
potential problem.
If the master observes that some player receives all
messages in the second round of epoch $k$
(Algorithm~\ref{alg:master},
line~\ref{line:checkCommit}), it concludes that some
byzantine player \emph{may} have committed this epoch.
Therefore, in this case, the master includes epoch
$k$ in the closing state.
Since the private keys of byzantine players are
unavailable to the master, it signs the epoch with its
own private key, and sends it to all players in the new
configuration (committee) as the opening state.
A player that sees an epoch with the master's signature
refers to it as if it is signed by all players.
(Recall that the master is a trusted entity,
emulated by a BFT protocol.)

\input{sections/masterCode.tex}

\subsection{Protocol analysis}
\label{sub:proof}

To prove that our protocol is correct we need to show
that (1) it is safe and fair in case all the rational
players follow the protocol and there are at most $f <
n/2$ byzantine players (Section~\ref{subCorrect}), and
(2) following the protocol is an equilibrium for
all rational players (Section~\ref{subRationality}).
We refer to a player that follows the protocol as
\emph{follower}.

\subsubsection{Safety and fairness}
\label{subCorrect}

\paragraph{Safety.}
First we show that if non-byzantine players follow the protocol,
then there are always at lest $f+1$ followers on the
committee.
Initially, the committee consists of $n \geq 2f+3$
players, of which at least $f+1$ members follow the
protocol.
Now since the $detect()$ operation of the DA2A
abstraction never returns members that follow the
protocol in case there are at least $f+3$ such members,
we get that the master never removes a player that
follows the protocol, and thus there are always at
least $f+1$ followers on the committee.

Now we show that if one player commits a sequence of
transactions in epoch $k$, no other committee member commits a
different sequence of transactions in epoch $k$.
Note that in order to commit a sequence of transactions
$st$, players must have proof that $st$ is allowed
to be committed.
One option for such proof is to have a $\langle
\mathsf{newConfig}, *, k, h(st) \rangle$ message from the
master, and another option is to have round 3
messages from $f+1$ committee members, each of which
contains a hash of $st$.

First note that every two members that commit epoch $k$
after receiving $\mathsf{newConfig}$ from the master
commit the same sequence of transactions because the master does
not send different messages in the same epoch.
Second, since all followers sends the same hash
of transactions to all committee members in round
$2$, all followers that send round 3 messages include the same
hash therein.
And since players that commit with the second option must have in
the proof at least one message from a follower,
all players that commit with the second option commit the same
sequence.
It remains to show that members that commit with the first
option and members that commit with the second one
commit the same sequence of transactions.
Let $p_j$ be a committee member that commits a sequence
of transactions $st$ with the second option.
Since $p_j$ receives $f+1$ messages in round 3, then it
received a round 3 message from at least one follower $p_f$.
Moreover, $p_f$ sent the round 3 message to $p_j$ before it 
received $\mathsf{reconfig}$ from the master.
In addition, since $p_f$ sent a round 3 message, then it
received round 2 messages that contains the hash of $st$
from all committee members before it received
$\mathsf{reconfig}$ from the master.
Therefore, $p_f$ includes all the rounds 2 messages that
it received in the $\mathsf{status}$ reply to the master,
and thus the master includes $st$ in the closing state, and
sends $\langle \mathsf{newConfig}, *, k, h(st) \rangle$ to the
new committee.
Hence, all members that commit with the first option
commit $st$ as well.

\paragraph{Fairness.}
We need to show that every committed epoch contains
transactions of all committee members.
First, note that the hash of transactions each player
sends in round 2 contains its own transaction.
Second, a player commits a sequence of transactions $st$
only
if some player receives
the hash of $st$ from all committee members in the second
round.  
Therefore, a player commits a sequence of
transactions only if it contains transactions of all
committee members.

\subsubsection{Rationality}
\label{subRationality}

We now show that following the protocol is an equilibrium
for all rational players. 
We first discuss committee players, and then the auditors that
emulate the master.

\paragraph{Committee players.}

First, players cannot increase their ratio in the
ledger by simply submitting more transactions than their
allocated QoS, because the QoS allocation is known to all, and
the excessive messages will be ignored. 
In addition, since a round 2 message is required from all
committee members in order for an epoch to be committed, and
since no committee member will sign a hash on a sequence that
excludes its transaction, we get that a player on the committee
cannot be excluded from a committed epoch.
Therefore, players cannot increase their ratio
in the ledger by (any) active deviation from the
protocol.
Moreover, since the master may punish them for an active
deviation by reducing their ratio (or removing them from
the committee), following the protocol is a better
strategy for them than any active deviation.

As for passive deviations, a possible strategy for a
rational player $p_i$ is to try to ``frame'' another
player $p_j$ and get it removed by the master, in which
case $p_i$'s ratio in the ledger will grow.
It can try to do this by not sending messages to $p_j$ or
by lying about not delivering $p_j$'s messages.
%
Now recall that our DA2A abstraction never wrongly accuses players
for passive deviation as long as there are at most $f+1$
deviating players.
Since we assume that players do
not collude, even if $p_i$ deviates, there are at most $f+1$
such players ($p_i$ plus $f$ byzantine players).
Therefore, it is impossible for $p_i$ to increase his ratio by 
passive deviation.
Moreover, since we assume that for a fixed ratio players prefer long
ledgers, sending protocol messages as fast as possible is an
equilibrium.


Finally, we argue that a rational player will ignore
reconfiguration change requests that do not have proofs.
This is because all players need to move to the same new
configuration in order to commit transactions, and so accepting
invalid remove will stall the protocol.

\paragraph{Master auditors.}
Consider first auditors who are not players. 
In case there is progress, the utility function of
the auditors is the number of players on the committee.
Otherwise, it is zero.
Therefore, auditors have no incentive to remove players
as long as there is a progress, and it is in their best
interest to detect and remove deviating players when
they stall the sequencing protocol. 

Second, consider auditors who also act as players.
Again, the only possible strategy for an auditor $p_i$ that
is also a player to increase its utility function is to
try to remove another player $p_j$ from the committee.
However, since players will not remove $p_j$
without a valid proof from the DA2A, $p_i$ cannot cause $p_i$'s
removal even if $f$ byzantine auditors also try to remove $p_i$.

\input{sections/evaluation.tex}

\section{Related Work}
\label{sec:related}


\paragraph{Fairness and rationality.}

Our work is indebted to recent works that combine
game theory and distributed
systems~\cite{
abraham2013distributed, abraham2011distributed,
aiyer2005bar, li2006bar, Moscibroda, feigenbaum2000sharing,
feldman2006free, srinivasan2003cooperation} to implement
different cooperative services.
In particular, we adopt a BAR-like model~\cite{aiyer2005bar,
li2006bar, Moscibroda}.
As in previous works on BAR fault
tolerance~\cite{aiyer2005bar, li2006bar}, we assume
non-colluding rational players, whereas colluding players
are deemed byzantine.
As in~\cite{Moscibroda}, , we do not assume
altruistic players -- all non-byzantine players are
rational in our model.

Practical byzantine fault tolerant consensus
protocols~\cite{golan2018sbft, canetti1993fast, abd2005fault,
yin2003separating, castro1999practical, kotla2007zyzzyva,
amir2006scaling, martin2006fast, li2007beyond,
amir2007customizable, clement2009making, 
veronese2009spin, amir2011prime, miller2016honey,
liu2015xft, duan2014bchain} have been studied for more than
two decades, but to the best of our knowledge, none deals with
rational players. 
Moreover, we are only familiar with two previous works that
consider some notion of fairness:
Prime~\cite{amir2011prime} and
Honeybadger~\cite{miller2016honey}.

One of the important insights in Prime~\cite{amir2011prime} is
that the freedom of the leader to propose transactions
must be restricted and verified by other participants.
To this end, Prime extends PBFT~\cite{castro1999practical}
with three additional all-to-all communication rounds at
the beginning, in which participants distribute among them
self transactions they wish to append to the ledger.
The leader proposes in round 4 a batch of transactions
that includes all sets of transactions it gets in round 3 from $2f+1$
participants. 
Since each transaction proposed by some participant is
passed to the leader by at least $2f+1$ participants, its
participant may expect its transaction to be proposed.
In case a participant send a request and the leader does
not propose it for some time $T$, the participant votes
to replace the leader.
As a result, Prime guarantees that during synchronous
periods every transaction is committed in a bounded time
$T$.

Similarly to FairLedger, Prime uses batching to commit
transactions of different participants atomically
together, and uses a PKI to ensure fairness and provide
proofs that the batches are valid.
However, their fairness guarantee is weaker than ours. 
Since the first three rounds are asynchronous (i.e.,
participants do not wait to hear from all, but rather
echo messages as soon as they receive them), there is no
bound on the ratio of transactions issued by different
participants that are committed during $T$.
More importantly, Prime assumes that all non-byzantine
participants follow the protocol, and we do not see a
simple way to adjust to overcome rational behavior.
For example, there is no incentive for participants to
echo transactions issued by other participants in the
first three rounds; to the contrary -- the less they
echo, the less transactions from other participants will
be proposed by the leader.

Honeybadger~\cite{miller2016honey} is a recent protocol
for permissioned blockchians, which is built on top of an
optimization of the atomic broadcast algorithm by
Cachin et al.~\cite{cachin2001secure}.
It works under fully asynchronous assumptions and provides
probabilistic guarantees.
Honeybadger assumes a model with $n$ servers and
infinitely many clients.
In brief, clients submit transactions to all the servers,
and servers agree on their order in epochs.
In each epoch, participants pick a batch of transactions
(previously submitted to them by clients) and use an efficient
variation of Bracha's reliable
broadcast~\cite{bracha1987asynchronous} to disseminate the
batches.
Then, participants use a randomized binary consensus
algorithm by Ben-Or et al.~\cite{ben1994asynchronous}
for every batch to agree whether or not to include it in the epoch.

Similarly to FairLedger, they use epochs to batch
transactions proposed by different players, and commit them atomically
together.
Their (probabilistic) fairness guarantee
is stronger than the one in Prime: they bound the number of
epochs (and accordingly the number of transactions) that can be
committed before any transaction that is successfully
submitted to $n-f$ servers.
However, if we adapt their protocol to our model where we do not
consider clients and require fairness among players, we
observe that their guarantee is weaker than ours:
Since communication is asynchronous, it may take arbitrarily
long for a transaction by player $p_i$ to get (be submitted) to
$n-f$ players, and in the meantime, other players may
commit an unbounded number of transactions.
In addition, their protocol uses building blocks (e.g.,
Bracha's broadcast~\cite{bracha1987asynchronous} and
Ben-Or et al.~\cite{ben1994asynchronous} randomized
consensus) that are not designed to deal with rational behavior.
Moreover, rational players that wish to increase their
ratio in the ledger will not include transactions issued
by other players in their batches.

Finally, it worth noting that both Prime and
Honeybadger are much more complex than FairLedger.
Prime's description in~\cite{amir2011prime} is spread
over more than 6 pages, and the reader is referred to
their full paper for more details.
Honeybadger combines several building blocks (e.g., the atomic
broadcast by Cachin et al.~\cite{cachin2001secure}), each
of which is complex by itself.

\paragraph{BFT protocols and assumptions.}

The vast majority of the practical BFT
protocols~\cite{duan2014bchain, liu2015xft,
yin2003separating, kotla2007zyzzyva, amir2006scaling,
martin2006fast, li2007beyond, amir2007customizable},
staring with PBFT~\cite{castro1999practical} assume a
model with $n$ symmetric servers (participants) that
communicate via reliable eventually synchronous channels.
Therefore, they can tolerate at most $f <n/3$ byzantine
failures~\cite{fischer1986easy}, and cannot accurately
detect participants' passive deviations
(withholding a message or lying about not
receiving it);
intuitively, it is impossible to distinguish whether a
player maliciously withholds its
message or the message is just slow.
Since passively deviating
participants cannot be accurately detected, they cannot
be punished or removed, and thus byzantine
participants can forever degradate
performance~\cite{clement2009making}, and rational
behavior cannot be disincentivize.

We, in contrast, assume synchronous communication, which
together with the use of a PKI allows FairLedger to be
simple, tolerate almost any minority of byzantine failures,
guarantee fairness, detect passive as well as active
deviations, and penalize
deviating players.
FairLedger uses the synchrony bound only to detect and
remove byzantine players that prevent progress, allowing
it to be very long (even minutes) without hurting normal
case performance. 
To reduce the cost of using a PKI, FairLedger signs only
the hashes of the messages.
Moreover, in WAN networks the cost of PKI is reduced due to
longer channels delays. 

As illustrated by works on Prime~\cite{amir2011prime} and
Aardvark~\cite{clement2009making}
most BFT protocols are vulnerable to performance
degradation caused by byzantine participants.
To remedy this, Aardvark focuses on improving the worst
case scenario.
We, on the other hand, follow the approach taken in
Zyzzyva~\cite{kotla2007zyzzyva}, and optimize the
failure-free scenario. 
We take this approach because byzantine failures are rare
in financial settings, and one can expect break-ins to be
investigated remedied.

We implement FairLedger inside Iroha~\cite{iroha}, which
is part of the Hyperledger~\cite{hyperledger} project.
Specifically, we substitute the ledger protocol in Iroha, 
which was originally based on the BFT protocol in
BChain~\cite{duan2014bchain}, with FairLedger.
In brief, their protocol consists of a chain of $3f+1$
participants, where the first $f+1$ order transactions.
To deal with a passively deviating participant that
withholds messages in the chain, they transfer both the sender and the
receiver (although only one of them deviates from the protocol)
to the back of the chain, where they do not take part in
ordering transactions.
Similarly to FairLedger, they assume synchrony with
coarse time bounds and use it to detect passive deviations.
However, in contrast to FairLedger, they do no accurately detect
byzantine players and punish correct ones as well.
Moreover, since the head of the chain decides on the
transaction order, Iroha does not guarantee fairness.

\paragraph{Broadcast primitives.}

In order to detect passive deviation we define
DA2A, a new detectable all-to-all communication
abstraction.
Even though many practical byzantine
broadcasts~\cite{bracha1985asynchronous,
reiter1995rampart, cachin2001secure,
drabkin2006practical, cachin2005asynchronous,
fitzi2006optimally, cristian1986atomic} were proposed
in the past, DA2A is the first to extend its API with a
$detect()$ method, which accurately returns all
misbehaving players.

\section{Discussion}
\label{sec:discussion}
Blockchains are widely regarded as the trading technology of the future; industry leaders in finance, banking, manufacturing, technology, and more are dedicating significant efforts towards advancing this technology. The heart of a blockchain is a distributed shared ledger protocol.  In this paper, we developed FairLedger, a novel shared ledger protocol for the blockchain setting.  Our protocol features the first byzantine fault-tolerant consensus engine to ensure fairness when all players are rational. It is also simple to understand and implement.  We integrated our protocol into Hyperledger, a leading industry blockchain for business framework, and showed that it achieves superior performance to existing protocols therein. We further compared FairLedger to PBFT in a WAN setting, achieving better results in failure-free scenarios.

%% file: sections/playerCode1.tex
\begin{algorithm}
 \caption{FairLedger committee member pseudocode.}
 \label{alg:player1}
\begin{algorithmic}[1]
\small

\Local{}{}
 \State $\emph{committee} \triangleq $ a set of players
 \State $epoch \in \mathbb{N}$, initially, $0$ 
 \State $TX \triangleq$ queue of new transactions for
 append calls
 \State $ledger \triangleq$ sequence of triples $\langle
 TX, H, C \rangle$, where 
  $TX$ is a 
  \hspace*{4mm} sequence of transactions,
 $H$ is a set of 
 signed 
  hashes, and 
  \hspace*{4.2mm} $C$ is a set  of signed
 commit messages, 
 initially empty.
\EndLocal

\Statex
 \LeftComment{Sequencing protocol}

\State $t \gets TX.dequeue()$
\While{true}
\label{line:SSbegin}

	
	 \LeftComment{round 1}
	\State broadcast($\mathsf{\langle epoch, 1
	\rangle}, t$) to \emph{committee} 
	\label{line:r1}
	\State collect responses in $ledger[epoch].TX$ 
	\State wait until $|ledger[epoch].TX| = |\emph{committee}|$
	\State order $ledger[epoch].TX$ by some predefined
	function
	\State $H \gets \{hash(ledger[epoch].TX)\}$

	\LeftComment{round 2}
	\State broadcast($\mathsf{\langle epoch, 2
	\rangle},H$) to \emph{committee} 
	\State collect responses in
	$ledger[epoch].H$ 
	\State wait until $|ledger[epoch].H| = |\emph{committee} |$
	\If{not all the hashes in $ledger[epoch].H$ are the
	same}
	
		\State complain to the master
		\label{line:complain}
		\Comment{active deviation detected}
		\State wait for a $\mathsf{reconfig}$ message from the
		master
	
	\EndIf

	\LeftComment{round 3}	
	\State broadcast($\mathsf{\langle epoch, 3
	\rangle}, commit, H$) to $\emph{committee}$
	\State collect responses in $ledger[epoch].C$ 
	\State wait until $|ledger[epoch].C| = f+1$ 
	
	 
	\State $epoch \gets epoch+1$
	\Comment{move to next epoch}
	\State $t \gets TX.dequeue()$
	\label{line:moveEpoch} 

\EndWhile

\Statex

\LeftComment{Reconfiguration}

\Receiving{$\mathsf{reconfig}$ from the master}
\label{line:reconfigBegin}

	\State send $\langle\mathsf{status}, ledger\rangle$ to
	the master
	\State wait for $\langle \mathsf{newConfig}, c, e,
		\emph{Hs} \rangle$ message from the master
		
	\State $\emph{committee}  \gets c$
	\State $epoch \gets e$
	\If{$\emph{Hs} = \bot$}
	
		\State empty $ledger[e]$
		
	\Else
	
		\State $ledger[e].H \gets \emph{Hs}$
		\State $epoch \gets epoch+1$
		\State $t \gets TX.dequeue()$ 
	\EndIf
	
	\State continue
\EndReceiving
\label{line:reconfigEnd}

%
%
%
%
%
%
%
%
%


\end{algorithmic}
\end{algorithm}


%% file: sections/masterCode.tex
\begin{algorithm}[ht]
 \caption{Code for the master.}
 \label{alg:master}
\begin{algorithmic}[1]
\small

\Local{}{}
 \State $\emph{committee} \triangleq $ a set of players
\EndLocal

\ForAll{epoch $e$, in order,}
	\ForAll{instance $D$ of DA2A in $e$, in order,}
	\label{line:forD}

		\State wait until $\emph{current-time} -
		\emph{D.start-time} > 2\Delta$
		\State $S \gets D.detect()$
		\label{line:detect}
	
		\If{$S \neq \emptyset$}
		 \State \emph{reconfigure$(S)$}
		\EndIf

	\EndFor
\EndFor

\Statex

\Procedure{\emph{reconfigure}}{$S$}
\label{line:recB}

	\State $R \gets \emptyset$; $\emph{new-committee}
	\gets \emptyset$ 
	\State send $\langle \mathsf{reconfig}
	\rangle$ to \emph{committee} 
	\State wait $2\Delta$ time 
	\State $ \emph{committee} \gets \emph{new-committee}
	\setminus S$
	\If{$\exists \emph{Hs} \in R$ that
	contains $|\emph{committee}|$ same hashes}
	\label{line:checkCommit}
	
		\State send $\langle \mathsf{newConfig}, \emph{committee},
		e+1,
		\emph{Hs} \rangle$ to  \emph{committee}

	\Else
	
		\State send $\langle \mathsf{newConfig}, \emph{committee}, e,
		\bot \rangle$ to \emph{committee}
		\State go to in line~\ref{line:forD}
		\Comment{need to check this epoch again}

	\EndIf
	

\EndProcedure
\label{line:recE}

\Statex

\Receiving{$\langle \mathsf{state},ledger\rangle$ from
player $p_i$} 

	\State $\emph{new-committee} \gets
	\emph{new-committee} \cup \{p_i\}$
	\State $R \gets R \cup \{ledger[e].H\}$

\EndReceiving

\end{algorithmic}
\end{algorithm}

%% file: sections/evaluation.tex
\section{FairLedger implementations}
\label{sec:imp}
We implement FairLedger based on Iroha's framework, written in C++.  
We intend to contribute our code to Hyperledger. Therefore, we only change Iroha's consensus algorithm (called Sumeragi \cite{sumeragi}) with our sequencing protocol, while keeping other components almost untouched (e.g., cryptographic components, communication layer, and client API). This implementation is described in Section~\ref{sec-imp-hlimp}. 

In order to evaluate the FairLedger protocol itself, independently of the Hyperledger framework, we implement another version of FairLedger's sequencing protocol based on PBFT's code structure, written in C++ as well. This implementation is described in Section~\ref{sec-imp-saimp}.      

\subsection{Hyperledger implementation}
\label{sec-imp-hlimp}
The Hyperledger framework consists of two types of entities, \emph{participants} (committee members in our case) that run the protocol, and \emph{clients} that generate transactions and send them to participants for sequencing. 

The FairLedger protocol at each participant is orchestrated by a single thread, referred to as \emph{logic thread}. The logic thread receives transactions from clients as well as messages from other participants into a wait-free incoming event queue. The connections between clients and participants are implemented as GRPC sessions \cite{grpc} (internally using TCP) sending Protobuf messages \cite{protobuf}.
The logic thread maintains a map of epoch numbers to epoch states. An epoch state consists of verified events of that epoch, one event slot per participant.

\begin{figure}[H]
  \centering 
  \includegraphics[width=\linewidth]{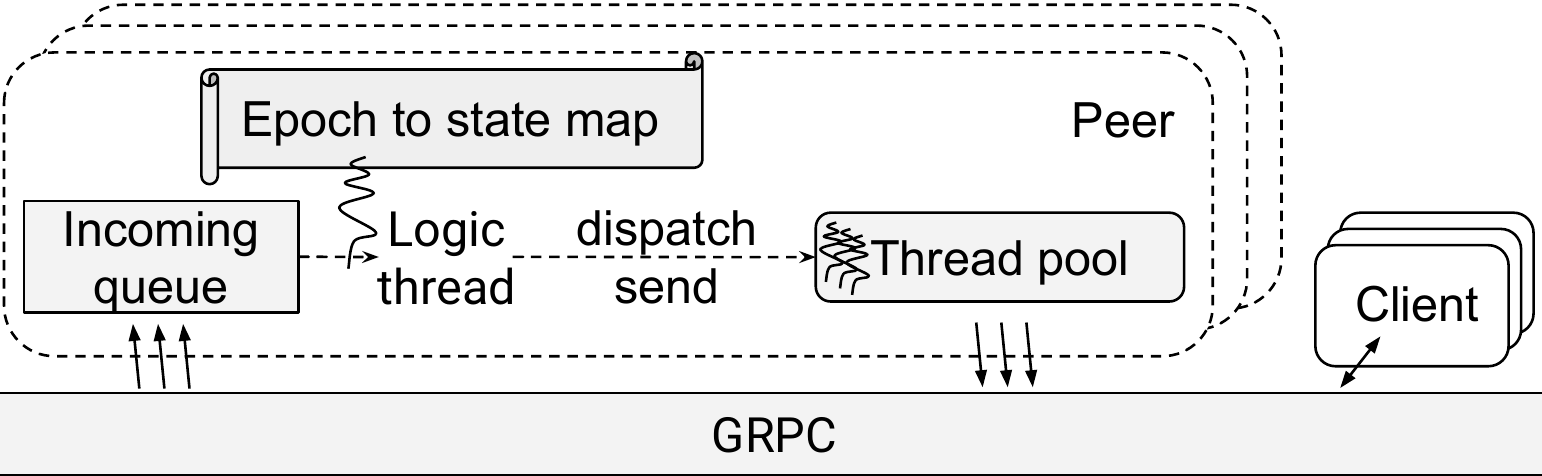}
  \caption{FairLedger implementation in Hyperledger.}
  \label{fig-fairDesign}
\end{figure}

Upon receiving a new message, the logic thread verifies it and decides based on the epoch state whether it needs to broadcast a message to other participants. Whenever broadcast is required, the logic thread creates and signs the new message, determines the set of its destinations (based on the epoch state), and creates send-message tasks, one per destination. These tasks are handed over to a work-stealing thread pool, in which each thread communicates with its destination over a GRPC connection (See Figure \ref{fig-fairDesign}).

Iroha is built in a modular fashion, which allows us to swap Sumeragi with FairLedger in a straightforward way.  
Our evaluation (in Section~\ref{sec-evaluation:iroha}) shows that additional Iroha components beyond the consensus engine adversely affect performance. 
Yet, these components are essential for Hyperledger. For example, Iroha supports multiple operating systems (including Android and iOS) and can be activated from java script code (via a web interface). Such features are essentials for client-facing systems like Iroha, and using standard libraries such as GRPC enables simple and clean development, which is less prone to bugs.

\subsection{Standalone implementation}
\label{sec-imp-saimp}

To eliminate the effect of the overhead induced the Hyperledger framework, we further evaluate the FairLedger protocol by itself, 
independently of the additional components. To this end, we employ the popular PBFT code~\cite{pbftCode} as our baseline. 
PBFT uses UDP channels, and is almost entirely self-contained, it depends only on one external library, for cryptographic functions. 

In this implementation of FairLedger, the logic thread directly communicates with clients and participants over UDP sockets. 
As in our Hyperledger implementation, the logic thread uses a map of epoch numbers to epoch states, and follows the same logic for  generating new messages.

Using UDP requires us to handle packet loss. We use a dedicated timer thread that wakes up periodically, (after a delay determined according to the line latency), verifies the progress of the minimal unfinished epoch, and requests missing messages from the minimal epoch if needed. 

\section{Evaluation}
\label{sec:eval} 
We now evaluate our FairLedger protocol using the two prototypes. The Hyperledger prototype is comparable to Iroha, 
and the standalone prototype is comparable to PBFT.
Section~\ref{sec-evaluation:env} describes the environment in which we
conduct our experiments and our test cases. 
Section~\ref{sec-evaluation:iroha} evaluates our
Hyperledger prototype, Section~\ref{sec-evaluation:pbft} compares FairLedger to PBFT, 
and Section~\ref{sec-evaluation:qos} evaluates performance under different QoS allocations.

\subsection{Experiment setup}
\label{sec-evaluation:env}

\paragraph{Configuration.} 
We conduct our experiments on Emulab~\cite{emulab}. 
We allocate 32 servers: 16 Emulab D710 machines for protocol participants, and 16 Emulab PC3000 machines for request-generating threads (clients).
Each D710 is a standard machine with a 2.4 GHz 64-bit Quad Core Xeon E5530 Nehalem processor, and 12 GB 1066 MHz DDR2 RAM. 
Each PC3000 is a single 3GHz processor machine with 2GB of RAM. 

Given that our system is intended for deployment over WAN among financial institutions, we configure the network latency among participants to 20ms. In Emulab, the communication takes place over a shared 1Gb LAN, denoted S-LAN. 
Each client is connected to a single (local) participant with a zero latency 1Gb LAN. 
In case clients need to communicate directly with remote participants (as they do in Iroha's design), they do so over S-LAN, i.e., with a latency penalty.  We benchmark the system at its throughput saturation point.

In our Hyperledger prototype evaluation, we use
 version v0.75.
Since in normal mode we assume no byzantine behavior, we configure Iroha with no faulty participants, so it signs each transaction once. The request-generating threads create transactions formatted according to Iroha's specification (given in Protobuf), which consists of a few hundreds of bytes of data.

In our standalone prototype evaluation, we create packets of a similar size, namely 512B of data, as this is the transaction size in our expected use case.

\paragraph{Test scenarios.}
\label{sec-evaluation:types}
We compare Iroha and PBFT to FairLedger's two operation modes -- the failure-free normal mode and the alert mode activated in case of attacks. 

We evaluate FairLedger's normal mode both using direct all-to-all and
using a single relay. We evaluate the alert mode both under attack of
a single byzantine participant, and without an attack. In the alert
mode we assume that $f$=1, and hence employ 3 relays. In the attack
scenario the byzantine participant remains undetectable by the
master. Specifically, one of the relays withholds messages that it
needs to send to one of the other relays.

In Section \ref{sec-evaluation:qos} we evaluate FairLedger with a slow participant that requires a lower QoS allocation.

\subsection{Hyperledger}
\label{sec-evaluation:iroha}
In order to deal with $f$ failures, FairLedger needs 2$f$+3
participants, and Iroha needs 3$f$+1. However, Iroha only uses 2$f$+1
signatures, and later broadcasts the committed requests to all 3$f$+1
participants. We chose to use only 2$f$+1 participants in favor of
Iroha, as it reduces Iroha's broadcast cost.
We scale our evaluation from 3 to 9 participants. Iroha's clients
perform asynchronous operations, and so the operation latency is
always zero. Hence, we focus this comparison on throughput.

\begin{figure}[H]      
  \centering 
  \includegraphics[width=\linewidth]{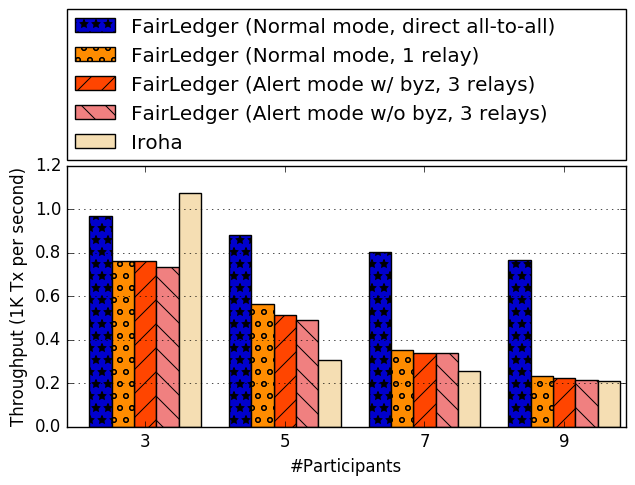}
  \caption{Throughput of FairLedger and Iroha over simulated WAN.}
 
   \label{fig-fairVsIroha}
\end{figure}

Figure~\ref{fig-fairVsIroha} compares the two modes of FairLedger with Iroha.
Results show that FairLedger's unrelayed normal mode has roughly the same throughput with 3 participants as Iroha, and much higher throughput  (up to 3.5x) than Iroha with more participants. In both algorithms, due to the usage of GRPC, the bottleneck is the broadcast. 
FairLedger commits more transactions per broadcast, since each epoch consists of one message from every participant, whereas 
Iroha pays the cost of broadcast for every client request. Therefore, Iroha suffers more as the broadcast cost increases (as we have more participants to send messages to).

FairLedger's relayed modes incur a 22\% reduction in throughput with 3 participants, and even more as the number of participants increases, because the relays worsen the bottleneck by issuing additional broadcast operations. Since in this implementation relaying is very costly, using relays in the normal mode is undesirable, and the performance reduction in alert mode is significant --  up to 66\%.
Note that a single router hampers FairLedger as much as three routers do. This is because of the protocol structure, where all nodes progress at the rate of the slowest one.  

Byzantine behavior slightly improves performance since withholding messages reduces the load on the relays. However, this effect is negligible.

\subsection{Standalone prototype}
\label{sec-evaluation:pbft}
\begin{figure}[H]      
  \centering 
  \includegraphics[width=\linewidth]{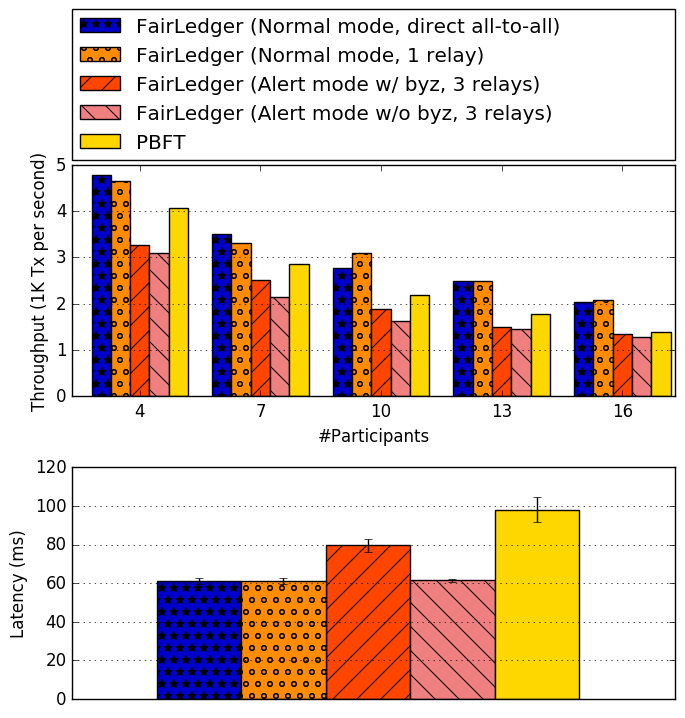}
  \caption{Throughput and latency of FairLedger and PBFT over simulated WAN.}
 
   \label{fig-fairVsPBFT}
\end{figure}

We evaluate our FairLedger prototype that is based on PBFT's code
structure.
We configure PBFT parameters in a way that maximizes PBFT's
throughput, enabling batching and enough outstanding client-requests
to saturate the system. We indeed achieve similar results to those
reported in recent work running PBFT over WAN~\cite{miller2016honey}.
In order to deal with $f$ failures, PBFT requires 3$f$+1
participants, so we run the evaluation with 4 to 16  participants.
Figure~\ref{fig-fairVsPBFT} shows the throughput and latency achieved
by the protocols.

First, we observe that the absolute throughput is 5x higher than with
Iroha.
This is thanks to PBFT's optimized bare-metal approach, which sacrifices modularity and maintainability for raw performance. We further see
 that FairLedger's normal mode has higher throughput than PBFT. This is because PBFT's clients are directed to a single participant (referred to as primary or leader), while FairLedger's clients address their nearest participant, distributing the load evenly among them.

FairLedger's alert mode with three relays reduces throughput by
30\%-40\% compared to the normal mode. Note that with 4 participants,
PBFT achieves about 25\% higher throughput than FairLedger's alert mode, but as the number of participants increases, the gap closes, reaching 9\% lower throughput than PBFT's with 16 participants.

The alert mode with one relay shows better performance than PBFT, and even better performance that FairLedger's normal mode in some cases. This happens because the single relay helps FairLedger overcome packet loss, while the cost of broadcasting a message over UDP is low. The packet loss is configured to be zero, i.e., there is no simulated loss, but our evaluation shows that it slightly increases with the number of participants due to load on the S-LAN. 

We measure latency below the saturation point. The results for all configuration sizes are similar, and so we depict in Figure 7 only the results with 10 nodes. Error bars depict the standard deviation.
The average latency of FairLedger clients in the normal mode is 64ms, which is close to the network latency of 3 rounds of 20ms. Indeed when  communicating over WAN, the performance penalty of signing and verifying signatures is negligible. 
PBFT's average latency is about 106ms, and consists of 3 PBFT rounds and 2 client-primary communication steps. 

The average latency of FairLedger's alert mode with a byzantine relay is 86ms, since it consists of 4 rounds of communication. The reason is that one participant is always one round behind the rest due to missing the byzantine participant's message. Since in the third round he require messages from $f$+1 participants (and not all of them), there is no need to wait for the lagging participant's round 3 message, and the epoch ends after 4 rounds. The latency of the alert modes without byzantine participants is 64ms, similarly to the normal mode.   

\subsection{QoS evaluation}
\label{sec-evaluation:qos}

We next show how QoS adjustment may help mitigate the throughput reduction due to a slow participant. 
We do so by testing the system in a relatively low rate, focusing solely on the effect of QoS adaptation.

\begin{figure}[H]      
  \centering 
  \includegraphics[width=\linewidth]{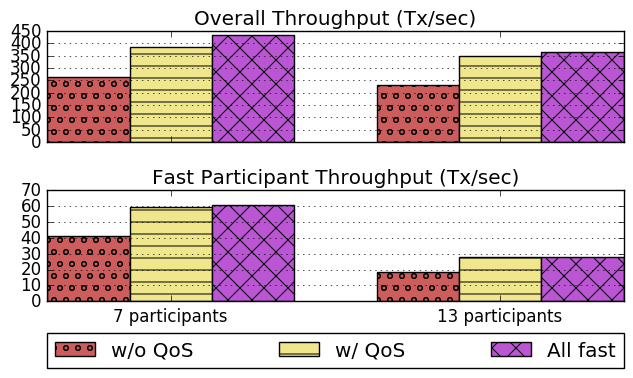}
  \caption{QoS of FairLedger.}
 
   \label{fig-fairQoS}
\end{figure}

We experiment with one slow participant that produces messages at half  the rate of other participants. Figure~\ref{fig-fairQoS} compares three scenarios: In the first, QoS is not enabled, and the slow participant dictates the rate of committed epochs. In the second the slow client's QoS is adjusted to 50\%. In the third scenario all participants are fast. 

Since the rate of the slow participant is half the rate of the fast participants, when QoS is adjusted, fast participants commit two transactions in every epoch,
while the slow participant commits one.
Results shows that when enabling QoS, the actual throughput that each fast participant receives is identical to the case in which all participants are fast. 

